\documentclass[aps,prd,floatfix,twocolumn,letterpaper,superscriptaddress]{revtex4}
\usepackage{amsmath,graphicx,bm,color}

\newcommand\ed{\varepsilon}
\newcommand\del{\partial}

\newcommand\hide[1]{}

\newcommand\fl[1]{\delta #1}
\newcommand\dv{\fl v}
\newcommand\dn{\fl n}
\newcommand\ded{\fl \ed}
\newcommand\dsn{\fl (s/n)}

\newcommand\pr[1]{\Delta^{#1}}
\newcommand\uv{u_\mathrm{v}}
\newcommand\ddd{\displaystyle}

\begin{document}

\title{Thermal chiral vortical and magnetic waves: new excitation modes in chiral fluids}

\author{Tigran Kalaydzhyan}\email{tigran@caltech.edu}
\affiliation{Department of Physics, University of Illinois, 845 W Taylor Street, Chicago, IL 60607, USA}
\affiliation{Jet Propulsion Laboratory, 4800 Oak Grove Dr, M/S 298, Pasadena, CA 91109, USA}

\author{Elena Murchikova}
\affiliation{TAPIR, California Institute of Technology, MC 350-17, Pasadena, CA 91125, USA}

\date{\today}

\begin{abstract}
In certain circumstances, chiral (parity-violating) medium can be described hydrodynamically as a chiral fluid with microscopic quantum anomalies. Possible examples of such systems include strongly coupled quark-gluon plasma, liquid helium $^3$He-A, neutron stars and the Early Universe. We study first-order hydrodynamics of a chiral fluid on a vortex background and in an external magnetic field. We show that there are two previously undiscovered modes describing heat waves propagating along the vortex and magnetic field. We call them the Thermal Chiral Vortical Wave and Thermal Chiral Magnetic Wave. We also identify known gapless excitations of density (chiral vortical and chiral magnetic waves) and transverse velocity (chiral Alfv\'en wave). We demonstrate that the velocity of the chiral vortical wave is zero, when the full hydrodynamic framework is applied, and hence the wave is absent and the excitation reduces to the charge diffusion mode. We also comment on the frame-dependent contributions to the obtained propagation velocities.
\end{abstract}

\maketitle

\section{Introduction}

Understanding of the transport phenomena in chiral systems progressed a lot in recent years.
It was mainly driven by the realization that microscopic quantum anomalies (such as the axial and mixed gauge-gravitational anomalies) can lead to macroscopic effects~\cite{Erdmenger:2008rm, Son:2009tf, Sadofyev:2010pr, Neiman:2010zi}, such as generation of unusual electric or axial currents and propagating excitations in rotating samples as well as in external electromagnetic fields. Interestingly, such phenomena are potentially observable in a wide span of physical situations, in, e.g., the Weyl/Dirac semimetals, strongly coupled quark-gluon plasma, in cold gases, superfluids and neutron stars, see Refs.~\cite{Miransky:2015ava, Kharzeev:2015kna} for a review. Hydrodynamics is a natural and widely used approach for the studies of anomalous dynamics in such many-body systems~\cite{Zakharov:2012vv}. The anomaly-driven effects appear already at the first order of derivative expansion. 

In this note we study first-order relativistic hydrodynamics with a single $U(1)$ charge and $U(1)^3$ triangle anomaly, in a background of a single vortex or external magnetic field. The hydrodynamic equations in their general form read
\begin{align}
\del_\mu T^{\mu\nu} = F^{\nu\alpha}J_\alpha, \qquad \del_\mu J^\mu = C\, E^\alpha B_\alpha\,, \label{nonconservation}
\end{align}
where $T^{\mu\nu}$ is the energy-momentum tensor, $J_\mu$ is the current, $E^\mu= F^{\mu\nu} u_\nu$ and $B^\mu = \frac{1}{2} \epsilon^{\mu\nu\alpha\beta}u_\nu F_{\alpha\beta}$ are the electric and magnetic fields in the fluid rest frame, $u^\mu$ is the fluid four-velocity, and $C$ is the axial anomaly coefficient. The choice of a single $U(1)$ is due to simplicity, and the generalization for several charges is straightforward.
In what follows, we will identify new propagating excitations, namely thermal chiral vortical and magnetic waves. Even though there were many instances when the vortical and magnetic waves were mentioned in the literature (see Ref.~\cite{Chernodub:2015gxa} and references therein), to our knowledge, there was no systematic study made in the hydrodynamic framework. The goal of our paper is to fill this gap and to identify the anomalous propagating modes.

\section{Chiral vortical wave}

In the absence of electromagnetic fields, the dynamics of a fluid is governed by the conservation laws
\begin{align}
\del_\mu T^{\mu\nu} = 0, \qquad \del_\mu J^\mu = 0\,. \label{conservation}
\end{align}
The velocity of the fluid in its own rest frame is $u^\mu = (1, 0, 0, 0)$ and, in general, $u^\mu = (1-\bm v^2)^{-1/2}(1, \bm v).$ We choose the Minkowski metric $\eta^{\mu\nu}= \mathrm{diag}(-1, 1, 1, 1)$ so $u^2 = -1$ and the transverse projector is defined as $\pr{\mu\nu}\equiv \eta^{\mu\nu} + u^\mu u^\nu$. In what follows we perform the calculations in the Landau frame (projections of first order terms in $T^{\mu\nu}$ and $J^{\mu}$ on $u^\mu$ vanish). This choice is due to the fact that we consider the case of zero average charge density, which is ill-defined in the Eckart frame.  Constitutive relations for the chiral fluid are the standard constitutive relations for ordinary fluid, but with an additional vortical term in the current~\cite{Son:2009tf},
\begin{align}
T^{\mu\nu} &= \ed u^\mu u^\nu + p \pr{\mu\nu} - \eta\pr{\mu\alpha}\pr{\nu\beta}\nonumber\\
& \times\left(\del_\alpha u_\beta + \del_\beta u_\alpha - \frac{2}{3}
\eta_{\alpha\beta}\del_\lambda u^\lambda\right) - \zeta \pr{\mu\nu} \del_\lambda u^\lambda,\\
J^\mu &= n u^\mu - \sigma T \pr{\mu\nu}\del_\nu (\mu/T) + \xi\, \omega^\mu\,,\label{Jmu}
\end{align}
where $n$ is the charge density, $\mu$ is the chemical potential, $\sigma$ is the conductivity, $\omega^\mu \equiv \frac{1}{2} \epsilon^{\mu\nu\alpha\beta} u_\nu \del_\alpha u_\beta$ is vorticity and $\xi$ is the chiral vortical coefficient~\cite{Neiman:2010zi}
\begin{align}
\xi = c_T {T^2} \left(1 - \frac{2 \mu n}{\ed + p}\right) + C \mu^2 \left(1 - \frac{2}{3}\frac{\mu n}{\ed + p}  \right)\,.
\end{align} 
The second term in $\xi$ is uniquely fixed by the requirement on the entropy current $s^\mu$ to satisfy $\del_\mu s^\mu \geq 0,$ see~\cite{Son:2009tf}. The prefactor for $T^2$ in the first term cannot be derived from hydrodynamics only~\cite{Neiman:2010zi} and is a manifestation of additional microscopic properties of the chiral degrees of freedom~\cite{Kalaydzhyan:2014bfa}. In a wide class of physical situations it is defined by the mixed gauge-gravitational anomaly~\cite{Volovik:2003fe, Landsteiner:2011cp, Jensen:2012kj}. In the simplest case, when the chirality is carried by non-interacting massless fermions of one chirality, flavor and color, $c_T=1/6$ and $C=1/(2\pi^2)$~\cite{Vilenkin:1979ui}.

Linearization  of equations (\ref{conservation}) around equilibrium leads to vortical term dropping out,
as a higher order correction, and recovery of the purely ordinary fluid dynamic equations studied in the literature~\cite{Minami:2009hn}.
In order to preserve the vortical term, we consider a vortex background given by the following velocity profile
\begin{align}
\uv^\mu = \left( 1, - \epsilon^{ijk}x_j \Omega_k\right) \equiv (1, \bm v_\mathrm{v})\,.\label{vortprof}
\end{align}
This corresponds to a constant external ${\bm \omega} = {\bm \Omega}$. It is important to emphasize that the rotation is driven by an external force and the velocity profile (\ref{vortprof}) is time-independent. In order to avoid relativistic corrections in $u^\mu$ coming from background rotation motion, we assume $R \ll 1/\Omega$, where   $R$ is the vortex radius (or a distance to the vortex center) and $\Omega = |\bm \Omega|$. The same will allow us to neglect contributions with $\omega^0 = \bm v \cdot (\bm \nabla \times \bm v) /2$ and its variations. Consider now fluctuations on top of this background,
\begin{align}
&\ed = \bar \ed + \ded, \, p = \bar p + \fl p,\, u^\mu= \uv^\mu+ \fl u^\mu,\, \fl u^\mu = (0, \fl {\bm v}),\nonumber \\
&n = \bar n + \dn,\, \mu = \bar \mu + \fl \mu, \, T= \bar T + \fl T,\, s = \bar s + \fl s\,.
\end{align}
We can treat the background values as constants, because the leading variations of the thermodynamic parameters due to the external rotation are at least quadratic in $\Omega$ \footnote{Odd powers of $\Omega$ would lead to the change of sign with the change in rotation direction}, e.g., $\Delta \bar \ed / \bar \ed \sim (\Omega R)^2 \ll 1$. At the same time, we allow the variations of the velocity $\fl \bm{v}$ to be the same order as the velocity $\bm{v}$, so they play the same role in the power counting. Substituting the variations into the conservations laws (\ref{conservation}), we get
\begin{align}
&\del_t \ded + \bar w \bm\nabla\cdot \fl\bm v = 0,\label{linhydro1}\\
&\bar w \del_t \fl\bm v = - \bm\nabla\fl p + \eta \nabla^2 \fl \bm v + (\zeta + \eta/3) \bm\nabla (\bm\nabla \cdot \fl \bm v)\label{linhydro2},\\
&\del_t \dn  + \bar n \bm\nabla\cdot \fl\bm v - \sigma T \del_\mu \pr{\mu\nu} \del_\nu(\mu / T) \nonumber\\
&~~+2 c_T{\bar T} \left( 1-\frac{2 \bar{\mu} \bar{n}}{\bar{w}} \right) \bm \Omega \cdot \bm \nabla \fl T + 2 C \bar \mu \bm \Omega \cdot \bm \nabla\fl\mu \nonumber\\
&~~-\frac{2 c_T \bar T^2 \bar{\mu} \bar{n}}{\bar w^2}\bm \Omega \cdot \bm \nabla \left( \frac{\bar w}{\bar \mu}\fl \mu + \frac{\bar w}{\bar n} \fl n - \ded - \fl p\right)\nonumber\\
&~~-\frac{2 C \bar \mu^3 \bar n}{3 \bar w^2}\bm \Omega \cdot \bm \nabla \left( \frac{3\bar w}{\bar \mu}\fl \mu + \frac{\bar w}{\bar n} \fl n
- \ded - \fl p\right) = 0\,,\label{linhydro3}
\end{align}
where  $\bar w \equiv \bar \ed + \bar p$ is the enthalpy density. Here we also used the identity $\bm\nabla \cdot \bm v_\mathrm{v}=0$ and dropped $\bm{v}\cdot\bm{\nabla}\fl X$ terms. 

Let us now perform the Fourier transform of
the hydrodynamic equations with the following convention
\begin{equation}
\Phi(t, \bm r) = \int_{-\infty}^{\infty} d\omega \int_{-\infty}^{\infty} d\bm{k}\, e^{-i\omega t + i \bm k \cdot \bm r}\, \Phi(\omega, \bm{k})\,.
\end{equation}
Note that $\bm \Omega$ enters equations in a scalar product with $\bm k$, and, therefore, components 
of $\bm k$ perpendicular to $\Omega$ do not contribute to anomalous terms.  
Therefore, to make the contribution of the anomaly more pronounced and for the additional simplicity, we assume that the wave-vector $\bm k$ is oriented along $\bm\Omega$. For some of the effects related to the general orientation $\bm k$, see Ref.~\cite{Abbasi:2015saa}. We further decompose $\bm v = \bm v_{\scriptscriptstyle\parallel} + \bm v_{\scriptscriptstyle\perp}$ into a sum of components parallel and transverse to $\bm k$, respectively. After that the system of equations (\ref{linhydro1})-(\ref{linhydro3}) becomes
\begin{align}
& \omega \, \ded=  k \bar w  \, \dv_{\scriptscriptstyle\parallel}\,, \label{eq:ded}\\
& \omega \, \dv_{\scriptscriptstyle\parallel} =  k \fl p / \bar w - ik^2 \gamma_s \dv_{\scriptscriptstyle\parallel}\,,\label{eq:dv}\\
&\omega \fl\bm v_{\scriptscriptstyle\perp} + i \frac{\eta}{\bar w} k^2 \fl \bm v_{\scriptscriptstyle\perp} = 0\,,\label{transverse}\\
&\omega \left( \dn - \frac{\bar n}{\bar w}\ded\right) + i\sigma k^2 \fl \mu - i\sigma \frac{\bar\mu}{\bar T} k^2 \fl T \nonumber\\
&~~ - 2 c_T{\Omega \bar T} k \left( 1 - \frac{2 \bar \mu \bar n}{\bar w}\right) \fl T - 2 C \Omega \bar \mu k \fl \mu \nonumber\\\
&~~ +\frac{2 c_T \bar T^2 \bar\mu \Omega k}{\bar w^2}\left( \frac{\bar w \bar n}{\bar \mu}\fl \mu + \bar w \fl n- \bar n \ded - \bar n \fl p\right) \nonumber\\\
&~~ +\frac{2 C \bar \mu^3 \Omega k}{3 \bar w^2}\left( \frac{3\bar w \bar n}{\bar \mu}\fl \mu + \bar w \fl n- \bar n \ded - \bar n \fl p\right) = 0\,, \label{currentconservation}
\end{align}
where we denoted $\gamma_s\equiv (\zeta+4\eta/3)/\bar w$, $k = |\bm k|$, $\fl v_{\scriptscriptstyle\parallel} = |\fl \bm v_{\scriptscriptstyle\parallel}|$. 
Below we analyze the possible hydrodynamic modes arising from the
equations above.

\subsection*{Hydrodynamic modes in case $\bar n = 0$}

We have five linearized equations (equation (\ref{transverse}) is a vector equation) and expect to obtain five dispersion relations. Firstly, we notice that Eq.~(\ref{transverse}) is decoupled from all other equations, and we can write
\begin{align}
\omega = - i \frac{\eta}{\bar w} k^2\,,
\end{align}
which is the viscous relaxation mode for the transverse velocity fluctuations. Such $\omega$ trivially satisfies all other equations, because we can set all fluctuations but $\fl \bm v_{\scriptscriptstyle\perp}$ to zero.
Secondly, the condition $\bar n = 0$ decouples Eq.~(\ref{currentconservation}) from all other equations. Focusing on Eq.~(\ref{currentconservation}) with $\fl T = 0$, we obtain
\begin{align}
\omega = 2 \Omega \bar \mu \left( \frac{C}{\chi} -\frac{C \bar \mu^2}{3\bar w} - \frac{ c_T \bar T^2}{\bar w} \right) k 
 - i \frac{\sigma}{\chi} k^2 \label{CVW}\,,
\end{align}
where $\chi=(\del n / \del \mu)_{\mu=0}$ is the susceptibility. Formally, the dispersion relation above corresponds to the (``isothermal'') chiral vortical wave \cite{Jiang:2015cva}. Its speed of propagation is given by the factor in front of $k$ in the linear term. In our case it vanishes, since $\bar \mu = \bar n / \chi = 0$, and we reproduce the usual charge diffusion. It is important to notice that, since the term $n \bm v$ is missing in the definition of the current in Ref.~\cite{Jiang:2015cva}, their case is equivalent to $\bar n = 0$, and the wave obtained in \cite{Jiang:2015cva} should not actually exist. 

Finally, in case $\bar n = 0$ and arbitrary $\fl T$, we have a useful relation $s = c_v c_s^2$, where $c_v = \del \ed / \del T$ is the specific heat and $c_s = \sqrt{\del p / \del \ed}$ is the speed of sound. This allows us to rewrite equations (\ref{eq:ded}), (\ref{eq:dv}), (\ref{currentconservation}) in the following form
\begin{align}
& \omega \, c_v \fl T=  k \bar w  \, \dv_{\scriptscriptstyle\parallel}\,,\\
& \omega \, \dv_{\scriptscriptstyle\parallel} =  k c_v c_s^2 \fl T / \bar w - ik^2 \gamma_s \dv_{\scriptscriptstyle\parallel}\,,\\
& \left(\omega + i \frac{\sigma}{\chi} k^2\right) \dn = 2 c_T{\Omega \bar T} k \fl T \,,
\end{align}
where $\bar \mu$ was already put to zero. From here we obtain charge diffusion (\ref{CVW}) together with the usual sound modes,
\begin{align}
\omega = \pm c_s k - i \frac{\gamma_s}{2} k^2\,.
\end{align}
Above we reproduced the 5 familiar modes: two transverse relaxation modes, two sound modes and one thermal diffusion mode. Let us stress once again that the special case $\bar n = 0$ does not give rise to any new modes, contrary to \cite{Jiang:2015cva}.

\subsection*{Hydrodynamic modes in case $\bar n \neq 0$}

It is easy to see that the two transverse relaxation modes are still the same as in $\bar n = 0$ case, since $\fl \bm v_{\scriptscriptstyle\perp}$ are still decoupled from other fluctuations. Further, using the standard thermodynamic relations
\begin{align}
&d \ed = T ds + \mu d n,\quad d p = s d T + n d \mu,\\
&w = \ed + p = Ts + \mu n\,,
\end{align}
we rewrite the Eq.~(\ref{currentconservation}) in a more convenient form 
\begin{align}
&\frac{\bar n^2 \bar T}{\bar w} \dsn \left( \omega +\frac{2 c_T \Omega \bar T^2 \bar \mu k}{\bar w}
+ \frac{2 C \Omega \bar \mu^3 k}{3 \bar w}\right) \nonumber\\
&~~~~= i \frac{\sigma k^2}{\bar n \bar T} \left( \bar T \fl p - \bar w \fl T \right)
- 2 c_T{\bar T \Omega} k \left( 1-\frac{2 \bar \mu \bar n}{\bar w} \right) \fl T \nonumber\\
&~~~~~+ \left(\frac{2 C \bar \mu \bar s \bar T}{\bar n \bar w}-\frac{2 c_T \bar T^2}{\bar w}\right)\Omega k \left(\bar s \fl T- \fl p\right) \nonumber\\
&~~~~~- \left(\frac{2 C \bar \mu^3 \bar n}{3 \bar w^2}+\frac{2 c_T \bar T^2 \bar \mu \bar n}{\bar w^2} \right) \Omega k \fl p\,.\label{newn}
\end{align}
Let us consider first the case $\fl p = 0.$ We get
\begin{align}
&\omega = v_\Omega k - i D_T k^2,\qquad D_T \equiv \frac{\sigma \bar w^2}{\bar n^2 \bar T c_p}\,,\nonumber
\end{align}
\begin{align}
&v_\Omega \equiv 2\Omega \left( \frac{C \bar \mu \bar s^2 \bar T}{\bar n^2 c_p} - \frac{c_T \bar T}{\bar n c_p}(2 \bar w - 3 \bar \mu \bar n)\right) \nonumber\\
&~~~~~~~~~~- \frac{2 \bar \mu \Omega}{\bar w}\left( c_T \bar T^2 + \frac{C}{3} \bar \mu^2\right),  \label{trueCVW}
\end{align}
where $c_p \equiv n T (\del (s/n) / \del T)_p$ is the specific heat at constant pressure. The obtained dispersion relation corresponds to the usual thermal diffusion mode with diffusion coefficient $D_T,$ in the limit  $\Omega \rightarrow 0.$ However, in the presence of the vortex, this mode propagates with the speed $v_\Omega$ along the vortex. To our knowledge, it has never been discussed in the literature. We call this excitation the ``thermal chiral vortical wave". We also note that this excitation is different from the chiral heat wave described in Ref.~\cite{Chernodub:2015gxa}.

Physical meaning of the second term for the velocity (\ref{trueCVW}) is the subtraction of the difference between the no-drag and Landau frame velocities~\cite{Stephanov:2015roa}\footnote{Since the wave excitations are considered on a background of fluid resting in the direction parallel to $\bm k$, the wave velocity $v_{\mathrm{ND}}$ in the no-drag frame is given by the wave velocity $v_\mathrm{L}$ in the Landau frame with addition of the extra terms from Eq.~(\ref{frameV}).},
\begin{align}
u^\mu_{\mathrm{Landau}} = u^\mu_{\mathrm{no-drag}} + \frac{2 \mu}{w}\left(c_T T^2 + \frac{C}{3}\mu^2 \right) \Omega^\mu\nonumber\\
~~~~~~~~~~~~~~~~~~~~~~~~ + \frac{1}{2 w}\left(c_T T^2 + C \mu^2\right) B^\mu\,.\label{frameV}
\end{align}
The seeming sign mismatch between the corresponding terms in Eqs.~(\ref{trueCVW}) and (\ref{frameV}) comes from the fact that the velocities of the waves are defined with respect to the fluid flow. 

To find the other modes, one has to perform a more general calculation, similar to \cite{Minami:2009hn, Kovtun:2012rj}. 
In order to do so, we consider a system of coupled equations (\ref{eq:ded}), (\ref{eq:dv}), (\ref{newn}) and 
employ the following thermodynamic identities, to reduce the number of independent variables to three
\begin{align}
\fl p = & \,\frac{\bar w c_s^2}{\gamma}\left(\alpha_p \fl T + \frac{1}{\bar n} \dn\right),\\
\dsn = & \,\frac{1}{\gamma \bar n}\left(\frac{c_p}{\bar T} \fl T - \frac{\bar w c_s^2 \alpha_p}{\bar n} \dn\right),\\
\ded = & \,\frac{c_p}{\gamma}\fl T + \frac{\bar w}{\bar n}\left(1 - \frac{\bar T c_s^2 \alpha_p}{\gamma} \right) \dn,
\end{align}
where $c_s = \sqrt{(\del p / \del \ed)_{s/n}}$ is the speed of sound, $\alpha_p = -(1/n)(\del n / T)_p$ is the thermal expansivity at constant pressure and 
$\gamma = (\del (s/n) / \del T)_{p}/(\del (s/n) / \del T)_{n}$ is the ratio of specific heats, see Appendix A in \cite{Minami:2009hn} for useful relations between thermodynamic derivatives. 
With these substitutions made, the full system of equations can be written in the matrix form
\begin{align}
A  \left(\begin{array}{l}\dn \\ \fl v_{\scriptscriptstyle\parallel} \\ \fl T \end{array}\right) = 0\,,
\end{align}
where the matrix $A$ is defined as
\begin{widetext}
\begin{align}
A  = \left(
\begin{array}{c c c}
- \ddd\frac{\omega \bar T c_s^2 \alpha_p}{\gamma} - i \ddd\frac{\sigma \bar w c_s^2 k^2}{\bar n^2 \gamma} + \alpha_{11} \Omega k & 0 & \ddd\frac{\omega \bar n c_p}{\bar{w}\gamma} + i \ddd\frac{\sigma \bar w k^2}{\bar n \bar T}\left(1 - \ddd\frac{c_s^2 \alpha_p \bar T}{\gamma} \right) + \alpha_{13}\Omega k\\
\omega\left(1 - \ddd\frac{c_s^2 \alpha_p \bar T}{\gamma} \right) & - k\bar n & \omega \ddd\frac{c_p \bar n}{\gamma \bar w}\\
-\ddd\frac{c_s^2 k}{\bar n \gamma} & \omega + i \gamma_s k^2 & -\ddd\frac{c_s^2 \alpha_p}{\gamma}k
\end{array}
\right)\, .
\end{align}
\end{widetext}
Here the coefficients in front of $\Omega k$ are given by
\begin{align}
\alpha_{11}=& \, \frac{2 C c_s^2 \bar \mu^3}{3 \gamma \bar w}\left(1 - \alpha_p \bar T + \frac{3\bar s \bar T \bar w }{\bar \mu^2 \bar n^2} \right)\nonumber\\
&~~-\frac{2 c_T c_s^2\bar T^3}{\gamma \bar w \bar n}\left( \alpha_p \bar \mu^2 \bar n + \bar s \right),\\
\alpha_{13} = & \, - 2 c_T {\bar T} \left( 1-\frac{2 \bar T \bar s}{\bar w} \right) -\frac{2 C \bar \mu \bar s^2 \bar T}{\bar n \bar w} \nonumber\\
&~~+\frac{c_s^2 \alpha_p}{\gamma} \left( - \frac{2 c_T \bar T^3 \bar s}{\bar w} +
\frac{2 C \bar \mu^3 \bar n}{3 \bar w}+ \frac{2 C \bar \mu \bar s \bar T}{\bar n} \right)
\nonumber\\
&~~+\frac{2 c_p \bar n \bar \mu}{\gamma \bar w^2}\left( c_T \bar T^2 +\frac{C \bar \mu^2}{3} \right).
\end{align}
Three dispersion relations can be obtained from the condition $\mathrm{det} (A) = 0$. One of them is Eq.~(\ref{trueCVW}). The other two are straightforward to obtain, but they are too long to be written in this paper. The speed of propagation of these modes is a function of $\Omega$ and thermodynamic quantities. This mode describes a modified sound propagation. We can demonstrate it by taking the limit $\Omega \rightarrow 0,$ and recovering the familiar sound modes
\begin{align}
\omega = \pm c_s k - i \frac{\tilde \gamma_s}{2} k^2\,,
\end{align}
where 
\begin{align}
\tilde \gamma_s = \gamma_s + \frac{\sigma\bar w^2}{\bar n^2 \bar T c_p}\left[\gamma - 1 + c_s^2 \bar T \left( \frac{c_p}{\bar w} - 2\alpha_p\right) \right]
\end{align}
is the modified damping rate that coincides with the one obtained in \cite{Minami:2009hn}.

\section{Chiral magnetic wave}

It is easy to repeat the analysis for the case of the chiral magnetic wave~\cite{Kharzeev:2010gd}. To this end we add two terms, $\sigma F^{\mu\nu}u_\nu + \xi_B B^\mu$, to the right-hand side of  Eq.~(\ref{Jmu}), where  $\xi_B$ is the chiral magnetic coefficient in the Landau frame \cite{Neiman:2010zi}
\begin{align}
\xi_B = C \mu \left(1- \frac{1}{2}\frac{n \mu}{\ed + p} \right) - \frac{c_T}{2} \frac{n}{\ed + p} T^2\,.
\end{align}
\noindent We keep the electric field equal to zero, in order to satisfy the current conservation and keep the transverse velocity equations decoupled. Vorticity $\bm \omega$ is defined on the fluctuations of velocity (no vortex background). The Eq.~(\ref{linhydro2})
becomes 
\begin{align}
&\bar w \del_t \fl \bm v + \bm\nabla\fl p - \eta \nabla^2 \fl \bm v - (\zeta + \eta/3) \bm\nabla (\bm\nabla \cdot \fl \bm v)\nonumber\\
& ~~~~~~~~~~~~~~~= \left(\bar n \fl \bm v - \sigma T\bm \nabla (\mu / T) +\xi \bm \omega \right)\times \bm B \nonumber\\
&~~~~~~~~~~~~~~~~~~~+ \sigma \bm B (\bm B \cdot \fl \bm v) - \sigma B^2 \fl \bm v\,.
\end{align}
Let us focus first on the equations involving fluctuations in the transverse velocity $\fl \bm v_{\scriptscriptstyle\perp} = (\fl v_{\scriptscriptstyle\perp}^1, \fl v_{\scriptscriptstyle\perp}^2)$. Assuming $\bm k {\parallel} \bm B$ for the same reason as for $\bm k$ the vorticity in the previous section, we obtain
\begin{align}
&(- i \bar w \omega + \eta k^2 - i\xi k B/2 + \sigma B^2) \fl v_{\scriptscriptstyle\perp}^1 - \bar n B \fl v_{\scriptscriptstyle\perp}^2 = 0,\\
&(- i \bar w \omega + \eta k^2 - i\xi k B/2 + \sigma B^2) \fl v_{\scriptscriptstyle\perp}^2 + \bar n B \fl v_{\scriptscriptstyle\perp}^1 = 0,
\end{align} 
where $B = |\bm B|$. Condition on the zero determinant of the coefficient matrix gives us two modes,
\begin{align}
\omega = \pm \frac{B \bar n}{\bar w} - \frac{\xi B}{2 \bar w}k - i \left(\frac{\eta}{\bar w} k^2 + \frac{\sigma}{\bar w} B^2\right)\,,\label{Alfven}
\end{align}
where the first term corresponds to the Larmor frequency, and the second term describes a general form of the Chiral Alfv\'en Wave \cite{Yamamoto:2015ria} propagating with the speed $v_{\mathrm{CAW}}=\frac{\xi B}{2 \bar w}$ in the direction opposite to $\bm B$. It is important to notice that the original velocity $v_{\mathrm{CAW}} = c_T \bar T^2 B / (2 \bar w)$ obtained in Ref.~\cite{Yamamoto:2015ria} in the limit $T \gg \mu$ is nothing but the difference in velocities between the no-drag and Landau frames, see (\ref{frameV}).

Let us switch to the fluctuations of density/temperature along the magnetic field. With the new terms in the current, Eq.~(\ref{currentconservation}) becomes
\begin{align}
&\omega \left( \dn - \frac{\bar n}{\bar w}\ded\right) + i\sigma k^2 \fl \mu - i\sigma \frac{\bar\mu}{\bar T} k^2 \fl T - C B k \fl \mu \nonumber\\\
&~~ +\frac{c_T \bar T^2 B k}{2 \bar w^2}\left( \frac{2\bar w \bar n}{\bar T}\fl T + \bar w \fl n- \bar n \ded - \bar n \fl p\right)\nonumber\\\
&~~ +\frac{C \bar \mu^2 B k}{2 \bar w^2}\left( \frac{2\bar w \bar n}{\bar \mu}\fl \mu + \bar w \fl n- \bar n \ded - \bar n \fl p\right) = 0\,. \label{currentconservation1}
\end{align}
In the case $\bar n = 0$, we obtain the chiral magnetic wave dispersion relation,
\begin{align}
&\omega = v_\chi k - i D_\chi k^2\,,\nonumber\\
&v_\chi \equiv  \frac{C B}{\chi}-\frac{c_T \bar T^2 B}{2 \bar w},\quad D_\chi \equiv \frac{\sigma}{\chi}\,. \label{CMW}
\end{align}
We see that the first term in $v_\chi$, as well as the damping rate $D_\chi,$ reproduce the ones from Ref.~\cite{Kharzeev:2010gd}, while the second term in $v_\chi$ is due to the velocity difference (\ref{frameV}).
Let us now generalize the situation to $\bar n \neq 0$ and rewrite Eq.~(\ref{currentconservation1}) as
\begin{align}
&\frac{\bar n^2 \bar T}{\bar w} \dsn \left( \omega + C \bar \mu^2 \frac{ B k}{2 \bar w} + c_T \bar T^2 \frac{ B k}{2 \bar w}\right) \nonumber\\
&~~~ = i \frac{\sigma k^2}{\bar n \bar T} \left( \bar T \fl p - \bar w \fl T \right)+ \frac{C B \bar s \bar T k}{\bar n \bar w}\left(\bar s \fl T- \fl p\right)
\nonumber\\
&~~~ + \frac{c_T B \bar T k}{\bar w} \fl T- \left(\frac{C B\bar \mu^2 \bar n k}{2 \bar w^2} + \frac{c_T B\bar T^2 \bar n k}{2 \bar w^2}\right) \fl p\,.\label{newnB}
\end{align}
With $\fl p = 0$, this leads to a new ``thermal chiral magnetic wave'' mode
\begin{align}
&\omega = v_\chi^T k - i D_T k^2,\\
& v_\chi^T \equiv \frac{B \bar T}{\bar n^2 c_p}(c_T + C \bar s^2) - \frac{B}{2 \bar w}\left(c_T \bar T^2 + C \bar \mu^2 \right)\label{newCMW}
\end{align}
To our knowledge, this mode has never been discussed in the literature before. The second term for the velocity in (\ref{newCMW}) is, again, due to (\ref{frameV}).

\section{Conclusions and outlook}
The main result of this work is the discovery of new excitation modes in chiral fluid. They are described by wave solutions given in (\ref{trueCVW}) and (\ref{newCMW}), and correspond to heat waves propagating along the vortex or magnetic field due to the quantum anomalies. We reproduced a general form of the chiral Alfv\'en wave (\ref{Alfven}), the chiral magnetic wave (\ref{newCMW}) and demonstrated that the discussed in the literature chiral vortical wave \cite{Jiang:2015cva} is absent and simply reduces to the charge diffusion mode (\ref{CVW}). We identified frame-dependent contributions to the velocities of these waves. In particular, velocity of the chiral Alfv\'en wave from Ref.~\cite{Yamamoto:2015ria} is simply a difference between velocities of the no-drag and Landau frames, meaning that the wave should be absent in the no-drag frame.
As an outlook, we propose to study a more realistic situation with two charges: one vector and one axial. The equation on the current conservation in (\ref{conservation}) will be replaced by two equations, $\del_\mu J_V^\mu = 0$ and $\del_\mu J_A^\mu = 0$, and, therefore, we expect an additional hydrodynamic mode. The expressions for the chiral magnetic and vortical coefficients will slightly change~\cite{Neiman:2010zi, Kalaydzhyan:2011vx, Gahramanov:2012wz, Kalaydzhyan:2014bfa}. The proposed calculations will be straightforward and similar to the presented in this paper.

{\bf Acknowledgements.} We are very grateful to Misha Stephanov, Naoki Yamamoto, Ho-Ung Yee, Xin An and Maxim Chernodub for critical comments and suggestions. EM thanks John Estes and Hirosi Ooguri for the discussions of fluid gravity duality which prompted her thinking about chiral fluids. EM is particularly grateful to Dr. David and Barbara Groce for their kindness and continuous support. 
TK work was supported in part by the U.S. Department of Energy under Contract No. DE-FG0201ER41195.

\end{document}